\documentclass[aps,prl,showpacs,notitlepage,amsmath,amssymb,floatfix,twocolumn,superscriptaddress]{revtex4-2}

\usepackage{amsmath} 
\usepackage{amssymb}
\usepackage{hyperref}
\usepackage[normalem]{ulem}
\usepackage{mathtools} 

\DeclareMathAlphabet{\altmathcal}{OMS}{cmsy}{m}{n}
\usepackage{mathptmx}
\begin{document}

\title{Turbulence unsteadiness drives extreme clustering}
\author{F. Zapata}
\email[Corresponding author:]{fnzapata@df.uba.ar}
\affiliation{Universidad de Buenos Aires, Facultad de Ciencias Exactas y Naturales, Departamento de Física, Ciudad Universitaria, 1428 Buenos Aires, Argentina,}
\affiliation{CONICET - Universidad de Buenos Aires, Instituto de F\'{\i}sica Interdisciplinaria y Aplicada (INFINA), Ciudad Universitaria, 1428 Buenos Aires, Argentina.}
\author{S. Angriman}
\affiliation{Universidad de Buenos Aires, Facultad de Ciencias Exactas y Naturales, Departamento de Física, Ciudad Universitaria, 1428 Buenos Aires, Argentina,}
\affiliation{CONICET - Universidad de Buenos Aires, Instituto de F\'{\i}sica Interdisciplinaria y Aplicada (INFINA), Ciudad Universitaria, 1428 Buenos Aires, Argentina.}
\author{A. Ferran}
\affiliation{Université Grenoble Alpes, CNRS, Grenoble-INP, LEGI, F-38000 Grenoble, France}
\affiliation{Department of Mechanical Engineering, University of Washington, Seattle, Washington 98195-2600, USA}
\author{P. Cobelli}
\affiliation{Universidad de Buenos Aires, Facultad de Ciencias Exactas y Naturales, Departamento de Física, Ciudad Universitaria, 1428 Buenos Aires, Argentina,}
\affiliation{CONICET - Universidad de Buenos Aires, Instituto de F\'{\i}sica Interdisciplinaria y Aplicada (INFINA), Ciudad Universitaria, 1428 Buenos Aires, Argentina.}
\author{M. Obligado}
\affiliation{Université Grenoble Alpes, CNRS, Grenoble-INP, LEGI, F-38000 Grenoble, France}
\author{P.D. Mininni}
\affiliation{Universidad de Buenos Aires, Facultad de Ciencias Exactas y Naturales, Departamento de Física, Ciudad Universitaria, 1428 Buenos Aires, Argentina,}
\affiliation{CONICET - Universidad de Buenos Aires, Instituto de F\'{\i}sica Interdisciplinaria y Aplicada (INFINA), Ciudad Universitaria, 1428 Buenos Aires, Argentina.}

\begin{abstract} 
We show that the unsteadiness of turbulence has a drastic effect on turbulence parameters and in particle cluster formation. To this end we use direct numerical simulations of particle laden flows with a steady forcing that generates an unsteady large-scale flow. Particle clustering correlates with the instantaneous Taylor-based flow Reynolds number, and anti-correlates with its instantaneous turbulent energy dissipation constant. A dimensional argument for these correlations is presented. In natural flows, unsteadiness can result in extreme particle clustering, which is stronger than the clustering expected from averaged inertial turbulence effects. 
\end{abstract}

\maketitle 

% Introduction
One of the most counterintuitive effects of turbulence is that, when a fluid is loaded with inertial particles, the flow can segregate the particles instead of mixing them \cite{Brandt2022a,Obligado_2014}. This phenomenon, which results in the formation of clusters with enhanced particle density, is relevant in volcanic clouds \cite{Ongaro_2016, Ichihara_2023}, to explain cloud formation \cite{Shaw_2003} and electrification \cite{DiRenzo_2018}, in other geophysical and natural contexts \cite{Breier2018}, and for industrial applications. In homogeneous and isotropic turbulence, two mechanisms govern the formation of clusters: particles with small inertia are expelled out of vortices \cite{Wang_1993}, while particles with large inertia accumulate near points with zero net forces \cite{Coleman_2009}.

Turbulence is an out-of-equilibrium phenomenon that is often studied in the statistical steady state, i.e., when external forces and dissipation balance in such a way that the system has well defined time averages. However, in many natural and industrial systems this is not the case. Out-of-equilibrium systems can fluctuate randomly between two or more states, in such a way that time averages never converge \cite{Delatorre_2007}. Unsteadiness affects energy dissipation rates and the flow spectral properties \cite{Berti2023}. This in turn has an effect in the mixing and transport of particles. As an example, it has been reported that motile particles such as phytoplankton can change their direction of migration in response to  overturning events associated to the turbulent flow in which they move \cite{Sengupta2017}.

What is the effect of unsteadiness in passive particles' cluster formation? And is the formation and evolution of clusters in realistic flows driven by turbulence, by the flow unsteadiness, or by a combination of both? Here we show that the naturally occurring modulation of out-of-equilibrium systems in time has a drastic effect on turbulence parameters and in particle cluster formation. Moreover, we show that clustering correlates with a small time delay with the instantaneous Taylor-based Reynolds number of the flow, and anticorrelates with its instantaneous turbulent energy dissipation rate. Hysteresis is present in this process, indicating the particles preserve a memory of previous states. We present a dimensional argument that considers this phenomenon as a change in the particles' effective inertia (measured by the Stokes number) depending on the flow state. This result allows for estimation of turbulent parameters from particles measurements, and indicates that flow unsteadiness must be considered in the study of many multiphase flows.

% Methods
We performed direct numerical simulations (DNSs) of the incompressible Navier-Stokes equation 
\begin{equation}
    \partial_t \mathbf{u} + \mathbf{u} \cdot \boldsymbol{\nabla} \mathbf{u} = - \boldsymbol{\nabla} p + \nu \nabla^2 \mathbf{u} + \mathbf{F} ,
\end{equation}
where $\mathbf{u}$ is the solenoidal fluid velocity field ($\nabla \cdot \mathbf{u} = 0$), $p$ is the pressure per unit mass density, $\nu$ is the kinematic viscosity, and $\mathbf{F}$ is an external volumetric mechanical forcing. Equations are written in dimensionless units based on a unit length $L_0$ and a unit velocity $U_0$, and solved in a three-dimensional $2\pi L_0$-periodic cubic box with a parallel pseudo-spectral method using the GHOST code \cite{mininni2011, rosemberg2020}. Spatial resolutions of $N^3 = 512^3$, $768^3$, and $1024^3$ grid points were used, yielding increasingly larger Reynolds numbers with kinematic viscosities respectively of $\nu_{512} = 1.1 \times 10^{-3} L_0 U_0$, $\nu_{768} = 6.7\times10^{-4} L_0 U_0$, and $\nu_{1024} = 4.6\times10^{-4} L_0 U_0$. The external forcing ${\bf F}$ generates large-scale periodic counter-rotating columns (in the following abbreviated as CRC). It was used before to study unsteadiness in \cite{goto2015}, and is given by 
\begin{equation}
    \mathbf{F} =  F_0 [\sin(x) \cos(y) \hat{x} - \cos(x) \sin(y) \hat{y}]
\end{equation}
This forcing corresponds to an array of four counter-rotating vortices in the $xy$ plane, with translational symmetry in $z$. The first columnar vortex occupies the volume $[0, \pi L_0) \times [0, \pi L_0) \times [0, 2\pi L_0)$, and is separated from the others by two vertical shear layers in the middle of the domain, aligned respectively with the $xy$ and $xz$ planes.

We also performed DNSs of homogeneous and isotropic turbulence (HIT) with random forcing, to compare against the CRC runs, following the same procedures used for the CRC forcing and using $N^3 = 768^3$ and $1024^3$ grid points. In these simulations the flow was sustained using a forcing with fixed amplitude and random phases, which were slowly evolved in time with a correlation time of $0.5$ large-scale eddy turnover times to prevent the development of a mean flow. The forcing was applied at the lowest wave numbers, resulting in an integral length scale $L\approx 1.1 L_0$. The kinematic viscosities were $\nu_{768} = 3.1\times10^{-4} L_0 U_0$ and $\nu_{1024} = 2.1\times10^{-4} L_0 U_0$. A DNS similar to HIT, but with time dependent forcing amplitude to synthetically generate unsteadiness, is discussed in \cite{SM}. All simulations have $\kappa \eta > 1$, where $\kappa = N/3$ is the largest resolved wave number, $\eta = (\nu^3/\epsilon)^{1/4}$ is the dissipation scale, and $\epsilon$ is the energy dissipation rate. 

In all simulations we integrated a simple model of one way coupled and heavy point particles with equation of motion
\begin{equation}
    \dot{\mathbf{x}}_p = \mathbf{v}(t), ~~~\dot{\mathbf{v}} = \frac{1}{\tau_p} [\mathbf{u}(\mathbf{x}_p,t) - \mathbf{v}(t)],
\end{equation}
where $\mathbf{u}(\mathbf{x}_p,t)$ is the fluid velocity at the particle position $\mathbf{x}_p$ at time $t$, and $\mathbf{v}(t)$ and $\tau_p$ are respectively the particle velocity and the particle Stokes time. In each run different sets of particles were added, each with $10^6$ particles and with different values of $\tau_p$. The Stokes numbers of these sets, $\textrm{St} = \tau_p/\tau_\eta$ (where $\tau_\eta = (\nu/\epsilon)^{1/2}$ is the Kolmogorov dissipation time of the flow), were $\textrm{St} = 3$ and $8$ for all flows and all spatial resolutions considered. A third set with $\textrm{St} = 14$ was also evolved only in the simulations with $1024^3$ grid points. For the CRC runs, $\tau_\eta$ and $\textrm{St}$ are the time average over very long times.

\begin{figure} \centering
    \includegraphics[width=1\linewidth]{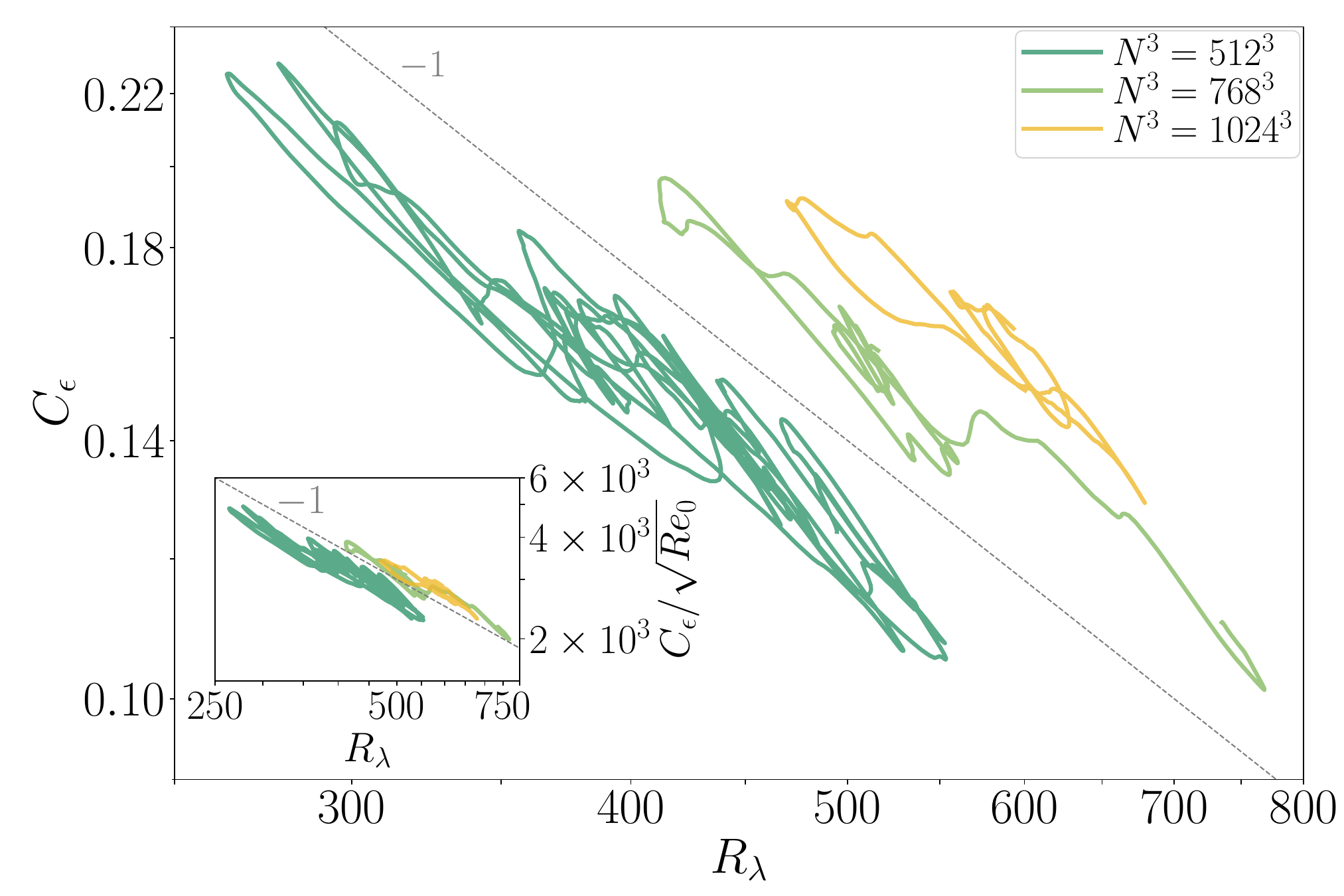} 
    \caption{Instantaneous value of $C_{\epsilon}$ as a function of $R_{\lambda}$ for the CRC flow at three different spatial resolutions. Each resolution corresponds to a different viscosity and to a different averaged Reynolds number. Inset: $C_{\epsilon}$ compensated by $\sqrt{\textrm{Re}_0}$ as a function of $R_\lambda$.}
    \label{ceps_relam}
\end{figure}

% Brief summary of the flow dynamics
The overall dynamics of the flows is as follows. While the HIT simulations display fluctuations in global quantities with a correlation time proportional to the integral turnover time, the CRC runs display distinct dynamics. Large excursions in the energy dissipation and other global quantities are observed, resulting from the flow transitioning from two states: one in which the large-scale columns can be clearly recognized (e.g., by direct inspection of the instantaneous spatial distribution of particles), and one in which the columns become unstable and the system displays a more homogeneous state. A movie of this time evolution can be seen in \cite{SM}.

% Results
An out-of-equilibrium dissipation law has been reported in a variety of unsteady turbulent flows \cite{Valente2014, vassilicos2023}, such that 
\begin{equation}
    C_{\epsilon} \sim \frac{\sqrt{\textrm{Re}_0}}{R_{\lambda}} ,
    \label{eq:ceps}
\end{equation}
where $\textrm{Re}_0 = u_0 l_0/\nu$ is a global Reynolds number based on the initial r.m.s.~flow velocity $u_0$ and the initial integral length scale $l_0$, and $R_{\lambda} = u(t) \lambda(t)/\nu$ is the local-in-time Reynolds number based on the instantaneous Taylor length scale $\lambda(t)$ and r.m.s.~turbulent velocity $u(t)$. $C_\epsilon$ is given by the energy dissipation rate as $\epsilon(t) = C_\epsilon u^3(t)/L(t)$, where $L(t)$ is the instantaneous flow integral scale. For a turbulent steady state (e.g., in HIT) this relation reduces to the well-known dissipation law $\epsilon = C_\epsilon u^3/L$ that states that the energy dissipation rate is governed by the large-scale energy flux towards smaller scales. In this sense, in unsteady flows $C_\epsilon(t)$ provides a measure of temporal scale-by-scale energy imbalance. 

In Fig.~\ref{ceps_relam} we see that the CRC flows display excursions compatible with Eq.~\eqref{eq:ceps} as $C_{\epsilon}$ is inversely proportional to $R_{\lambda}$. In the inset we show $C_{\epsilon}$ compensated by the square root of the reference Reynolds number $\textrm{Re}_0$; note how all curves from flows with different viscosities collapse. For a given viscosity (e.g., for $N^3 = 512^3$) $R_{\lambda}$ and $C_\epsilon$ change in time by factors of 2, with a typical time scale of the excursions of $10\langle T \rangle$ to $20\langle T \rangle$, where $\langle T \rangle$ is the mean large-scale eddy turnover time (see details below). These excursions, as well as their characteristic time scale, are much larger than those associated to the fluctuations in simulations of HIT (which take place in time scales of the order of the turnover time). In spite of these differences, the instantaneous energy spectrum of the CRC simulations still displays Kolmogorov scaling (not shown).

\begin{figure}[t] \centering
    \includegraphics[width=1\linewidth]{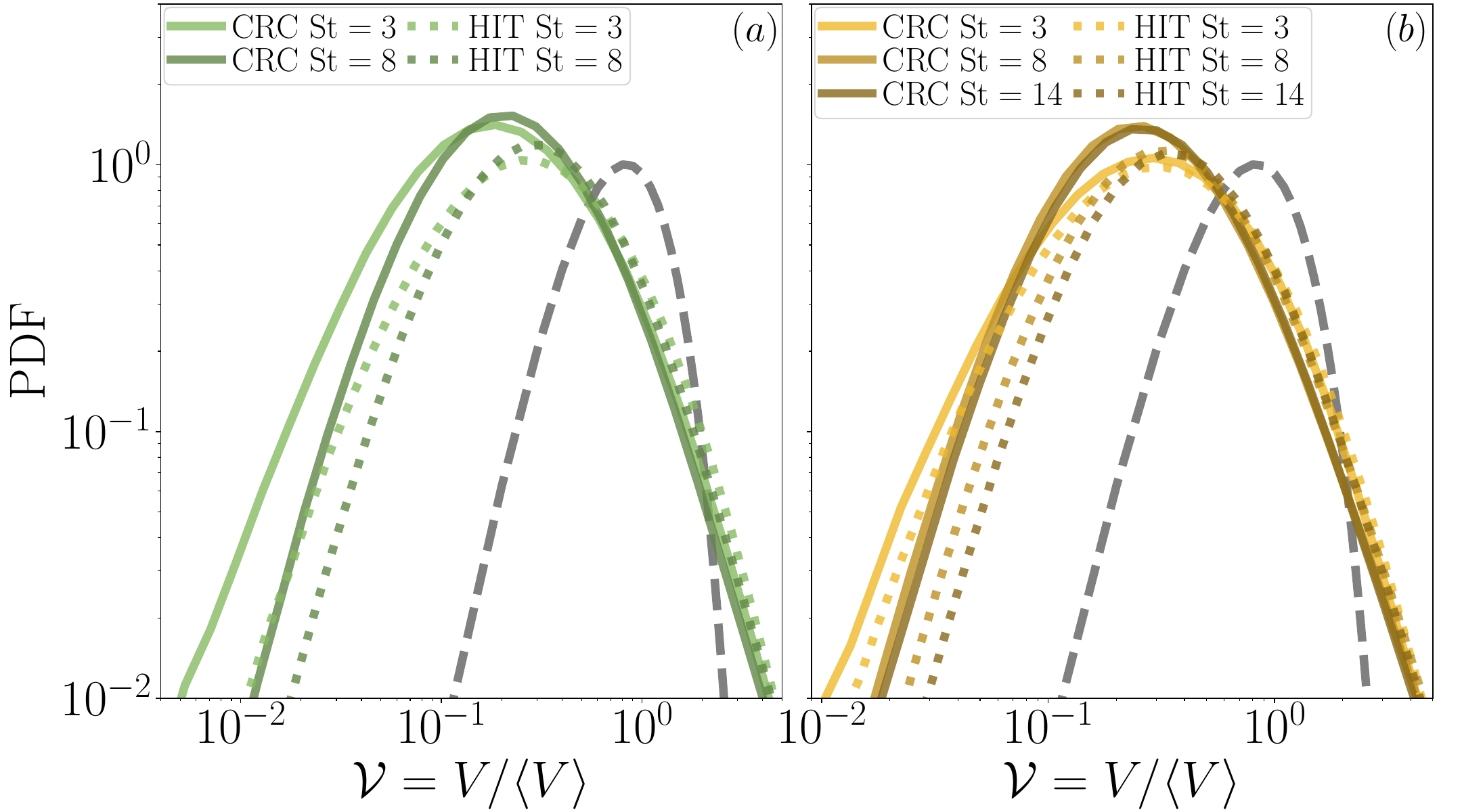} 
    \caption{Probability density functions (PDFs) of the normalized Vonoroï volumes $\altmathcal{V} = V/\left<V\right>$ of inertial particles for CRC and HIT simulations, for (a) $N^3 = 768^3$, and (b) $N^3 = 1024^3$. For CRC forcing, the PDFs are time-averaged over long times. A Random Poisson Process (RPP) is indicated by the dashed line.}
    \label{voropart}
\end{figure}

We want to know if these excursions in the dissipation and in the Taylor-based Reynolds number, similar to those reported in other unsteady turbulent flows \cite{Valente2014, goto2015, vassilicos2023}, affect particle cluster formation and time evolution. To estimate the amount of clustering in the different simulations we calculated the three dimensional Voronoï tessellation of the particles as a function of time.  Voronoï diagrams have proven to be a powerful tool to study particle clustering \cite{Monchaux2010, Monchaux2012}. The Vorono\"i cell associated to a given particle at a certain time is defined as the set of points closer to that particle than to any other particle. The volumes of the Voronoï cells $V$ were normalized by the mean volume of all cells $\left<V\right>$, to define normalized volumes $\altmathcal{V} = V/\left<V\right>$. Figure \ref{voropart} shows the time-averaged probability density functions (PDFs) of the normalized volumes, compared against the PDF resulting from a random Poisson process (herein RPP, corresponding to a homogeneous distribution of particles \cite{Uhlmann_2020}) which is shown as a reference. The stronger the tails of the PDFs compared against the RPP (i.e., the excess of probability for small $\altmathcal{V}$ corresponding to an excess of clusterized small volumes, or for large $\altmathcal{V}$ corresponding to large voids), and the larger the standard deviation of the PDFs, $\sigma_\altmathcal{V}$ (compared to the RPP which has a standard deviation of $\sigma_\textrm{RPP} \approx 0.42$), indicate enhanced clustering. Positions of the maxima also change as variables in the PDFs are normalized to obtain mean volumes occupied per particle.

\begin{figure}[t] \centering
    \includegraphics[width=1\linewidth]{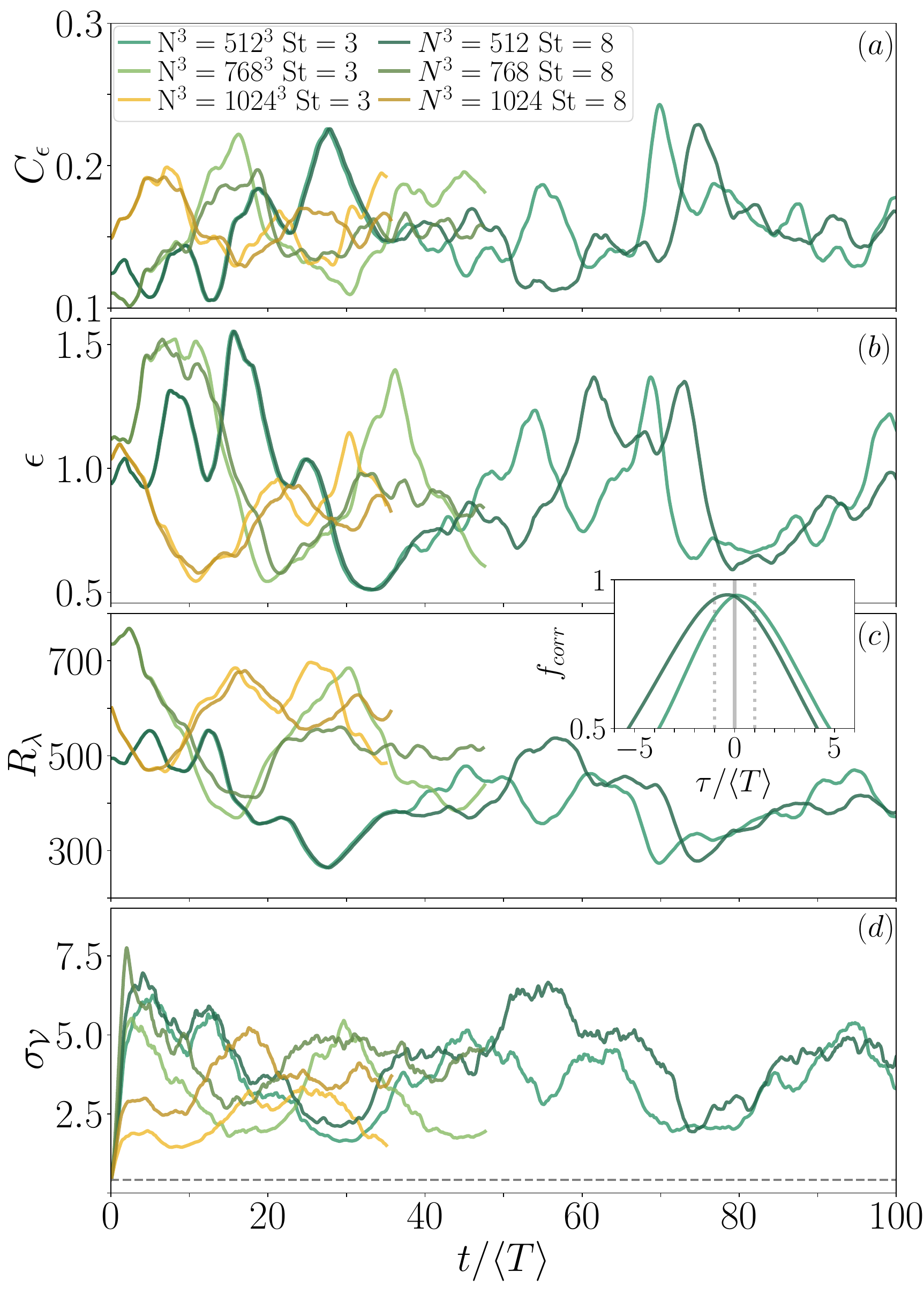} 
    \caption{Time series of (a) $C_{\epsilon}$, (b) $\epsilon$, (c) $R_{\lambda}$, and (d) $\sigma_{\altmathcal{V}}$ for CRC runs with $\textrm{St}=3$ and $8$. Time is normalized by the mean turnover time $\left<T\right>$, and $\sigma_\textrm{RPP}$ is indicated by the dotted grey line in panel (d). Inset: cross-correlation of $\sigma_\altmathcal{V}$ and $R_\lambda$ as a function of the time lag $\tau$.}
    \label{tsfcor}
\end{figure}

As shown in Fig.~\ref{voropart}, the CRC runs present stronger clustering than HIT, when comparing cases with the same Stokes number. As $\textrm{St}$ increases (at fixed spatial resolution) clustering diminishes, indicating that particles with more inertia cluster less both in CRC and HIT flows (results presented here are for $\textrm{St} \gtrsim 1$, as for $\textrm{St}\to 0$ and $\to \infty$ particles do not cluster; note that in the CRC flow for $\textrm{St}= 14$ clustering is similar to $\textrm{St}= 8$ and less than for 3). Differences between the CRC flow and HIT could in principle be associated with the presence of a large-scale flow in the CRC runs, but this is not sufficient to explain the observed enhancement in clustering. In other turbulent flows with a steady large-scale circulation, particle clustering was observed to be closer to that of HIT \cite{Angriman2022}. The reason for the stronger clustering here becomes more clear when inspecting the instantaneous PDFs of $\altmathcal{V}$. Note that all PDFs in Fig.~\ref{voropart} are averaged over a time window of $\approx 5 \langle T \rangle$, using $10^6$ Vorono\"i volumes in each snapshot with a cadence of at least $0.04 \langle T \rangle$. But while the PDFs of $\altmathcal{V}$ in HIT are stationary, the PDFs in the CRC runs are not (see a movie with the PDFs as a function of time in \cite{SM}).

\begin{figure}[t] \centering
    \includegraphics[width=1\linewidth]{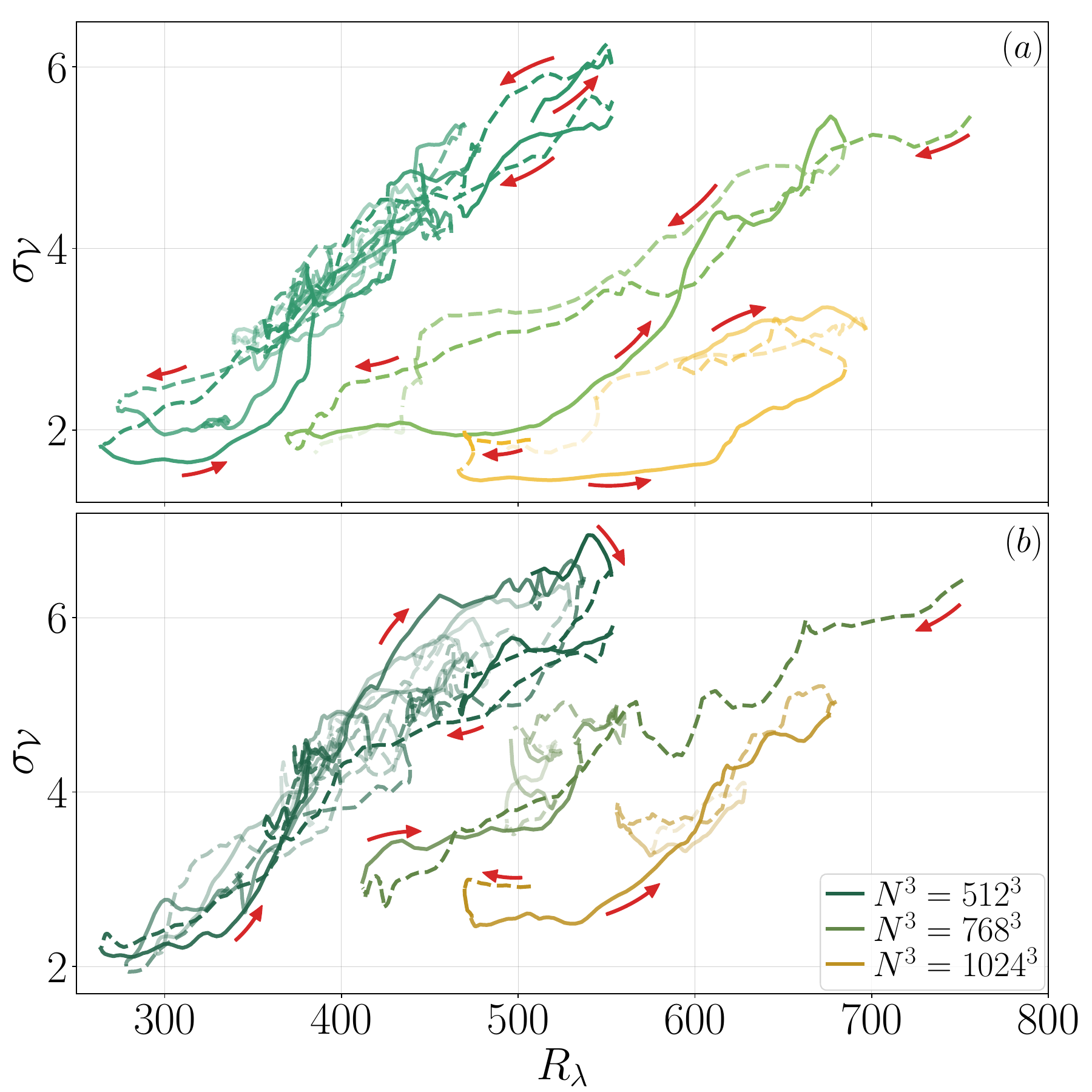} 
    \caption{Standard deviation of the Voronoï volumes, $\sigma_{\altmathcal{V}}$, as a function of $R_{\lambda}$ for CRC runs, differentiating branches in which $R_{\lambda}$ increases (solid lines) and decreases (dashed lines), for (a) $\textrm{St}=3$, and (b) $8$. Red arrows indicate the direction of time evolution.}
    \label{hyst}
\end{figure}

The instantaneous PDFs of the CRC flow vary significantly in time, and the time averaged PDFs showed in Fig.~\ref{voropart} alone are not representative of the actual level of particle clustering. To study how particle clustering is affected by the flow unsteadiness, we calculated the time evolution of $C_{\epsilon}$, $\epsilon$, $R_{\lambda}$, and the standard deviation of the Vorono\"i volumes, $\sigma_{\altmathcal{V}}$. Results are shown in Fig.~\ref{tsfcor}. Particles are injected when the flows are already in a fully developed turbulent regime, at a time arbitrarily labeled as $t=0$, and at random positions in space (note that at $t=0$, $\sigma_{\altmathcal{V}} = \sigma_{\textrm{RPP}}$ in all cases). After a short transient, particles form clusters as indicated by $\sigma_\altmathcal{V} > \sigma_\textrm{RPP}$. It has been reported before \cite{goto2015} that this flow displays irregular behavior of large-scale quantities. Our results for $C_\epsilon$ and $R_\lambda$ are compatible with this observation: both quantities display large excursions with a characteristic time scale much larger than the integral turnover time. Fluctuations in $C_{\epsilon}$ have a correlation time between $10\left<T\right>$ to $20\left<T\right>$, and result in fluctuations of the flow dissipation rate $\epsilon$ with a similar characteristic time. More surprising are the fluctuations in $\sigma_\altmathcal{V}$, which can vary between $\approx 2.5$ to values larger than 6, indicating strong variations in the level of particle clustering as the flow evolves.

Comparing the four quantities in Fig.~\ref{tsfcor} we see that when $C_{\epsilon}$ and $\epsilon$ display a local minimum, $R_{\lambda}$ and $\sigma_{\altmathcal{V}}$ display local maxima (i.e., turbulence becomes stronger and particles cluster more). To quantify the correlation between clustering and $R_{\lambda}$, the cross-correlation function $f_{corr}$ of $\sigma_\altmathcal{V}$ and $R_\lambda$ is shown in an inset in Fig.~\ref{tsfcor}, for the simulation with $512^5$ grid points as this simulation has the longest integration in time. The maximum cross-correlation is reached for time increments $|\tau|/\left<T\right> \lesssim 1$ (negative increments are associated to particles clustering during the growth of $R_\lambda$). Similar results are obtained when the cross-correlation is computed with $C_\epsilon$. Thus, $C_{\epsilon}$ or $R_{\lambda}$ and $\sigma_\altmathcal{V}$ are correlated with a time lag that is proportional to the large-scale eddy turnover time. The existence of this time lag suggests that the response of the particles to changes in flow properties may display hysteresis.

Figure \ref{hyst} shows $\sigma_\altmathcal{V}$ as a function of $R_{\lambda}$ in CRC runs, for particles with $\textrm{St} = 3$ and $8$, and for different viscosities and spatial resolutions. Time intervals with increasing $R_{\lambda}$ are marked with solid lines, while intervals with decreasing $R_{\lambda}$ are indicated with dashed lines. The figure indicates a Reynolds number dependence, and further confirms the correlation between these quantities and the existence of a time lag with a hysteresis cycle superposed over the strong fluctuations.

\begin{figure} \centering
    \includegraphics[width=1\linewidth]{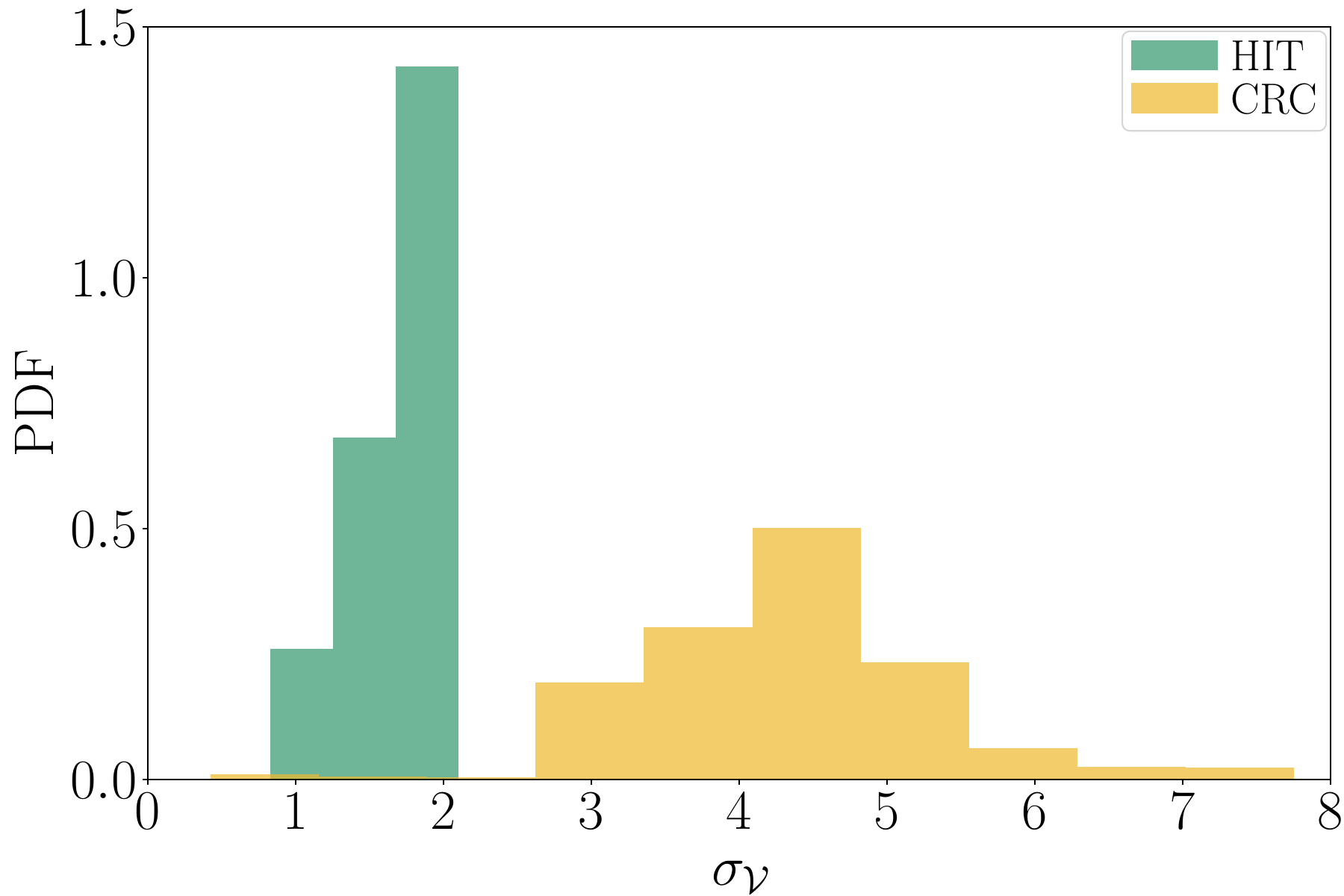} 
    \caption{PDFs of $\sigma_\altmathcal{V}$ for HIT and CRC runs with $N^3 = 768^3$ and for particles with $St = 8$. Note the larger values of $\sigma_\altmathcal{V}$ (i.e., enhanced clustering) in the CRC run, as well as the larger dispersion.}
    \label{pdfsigma}
\end{figure}

As previously mentioned, this level of particle clustering and its fluctuations are stronger than in HIT. Figure \ref{pdfsigma} compares the PDFs of $\sigma_\altmathcal{V}$ for CRC forcing and for HIT, in the case with $N^3=768^3$ and $\textrm{St} = 8$. The two flows display distinct values of $\sigma_\altmathcal{V}$. In HIT $\sigma_\altmathcal{V}$ takes values between $0.5$ and $1.8$, while in the CRC flow values go from $0.5$ to $7.4$, reaching its maximum value around $\approx 4.5$. Not only is $\sigma_\altmathcal{V}$ on average significantly larger in the CRC flow (i.e., particles cluster more), but the dispersion in the values of $\sigma_\altmathcal{V}$ is also much larger, including instances (albeit less probable) in which the clustering is similar to that found in HIT, as the ones captured by the left tail of the PDF. The dispersion in $\sigma_\altmathcal{V}$, and the correlation with $R_{\lambda}$, confirm that the extreme clustering observed in this flow is associated to its unsteady dynamics (see also \cite{SM}).

How does the out-of-equilibrium dynamics of the flow affect the particle behavior? We can consider first the situation in which turbulence is in a scale-by-scale steady state. Under these conditions, $\textrm{Re}_0 = u_0 l_0/\nu \approx u L/\nu$ (where $u = \langle u(t) \rangle$ is the time average of the r.m.s.~flow velocity, and $L = \langle L(t) \rangle$ is the averaged flow integral scale). The energy dissipation rate is also $\epsilon \approx u^3/L$. For a small spherical particle we can write the Stokes time as $\tau_p = 2 a^2 \rho_p /(9 \rho_f \nu)$ (where $a$ is the particle radius, $\rho_p$ is the particle density, and $\rho_f$ is the fluid density). Then $\textrm{St} = \tau_p/\tau_\eta = (2/9) (\rho_p/\rho_f) (a/L)^2 \textrm{Re}_0^{3/2}$. Thus, the Stokes number of the particles (and as a result, the sensitivity of the particles to flow fluctuations at different scales) is fixed given a particle radius, a mass density ratio, and a flow Reynolds number. However, when scale-by-scale steadiness is broken, we must use $\epsilon = C_\epsilon U^3/L$, which using Eq.~\eqref{eq:ceps} results in $\textrm{St} \sim (2/9) (\rho_p/\rho_f) (a/L)^2 \textrm{Re}_0^{7/4}/R_\lambda^{1/2}$. The sensitivity of the particles to flow fluctuations thus changes depending on the instantaneous dissipation (or the Taylor-based Reynolds number). Equivalently, we could interpret this expression as the effective ratio of the particle size to the flow scale being replaced by $a/(L R_\lambda^{1/4})$: the integral scale of the flow ``seen" by the particle depends on $R_\lambda$. Perhaps counter-intuitively, for larger $R_\lambda$ the particles become more sensitive to the larger scale eddies, which results in stronger clustering as can be seen in the movie in \cite{SM}. When $R_\lambda$ is smaller ($C_\epsilon$ larger) the effective Stokes number of the particles is larger and the spatial distribution of particles is more homogeneous, with clustering closer to that observed in inertial clustering in HIT. When $R_\lambda$ is larger ($C_\epsilon$ smaller), $\textrm{St}$ is smaller, and particles accumulate outside the large-scale columnar vortices in a phenomenon reminiscent of turbophoresis (i.e., expelled from the vortices \cite{Wang_1993}), resulting in extreme clusterization. As discussed previously, this argument can hold for $\textrm{St}$ larger than 1 but in its vicinity, as considered in our simulations.

Natural flows are unsteady: volcanic and other sources of particulate material pulsate and oscillate in time, convection in clouds is inhomogeneous, and atmospheric turbulence in general is bursty \cite{Rodriguez_Imazio_2023}. The results presented here show that flow unsteadiness, which in this case occurs naturally as a steady forcing is used, can drastically enhance particle clustering, well above what previous studies have reported for steady state homogeneous and isotropic turbulence. Thus this effect, which has been neglected so far, can change estimations of clustering and particle aggregation for many systems. As an example, estimations of collision frequencies between particles are proportional to $n^2 \sim \altmathcal{V}^{-2}$ (where $n$ is the particle density), and changes in the volume per particle $\altmathcal{V}$ affect the number of collisions. The results also open the door to the estimation of instantaneous turbulence parameters from direct observations of particle aggregation (albeit taking into account that the dependence on $\textrm{Re}_0$ could be different for different flows), and can be useful for the study of other out-of-equilibrium unsteady systems.

\begin{acknowledgments}
This work has been partially supported by ECOS-Sud Project No.~A18ST04. F.Z., S.A., P.C., and P.D.M.~acknowledge financial support from UBACyT Grant No.~20020170100508BA and PICT Grant No.~2018-4298. A.F. and M.O. acknowledge the LabEx Tec21 (Investissements d'Avenir - Grant Agreement $\#$ ANR-11-LABX-0030).
\end{acknowledgments}

\bibliographystyle{apsrev4-2}
\bibliography{ms}

\end{document}

\title{Supplemental Material: Turbulence unsteadiness drives extreme clustering}
\author{F. Zapata}
\affiliation{Universidad de Buenos Aires, Facultad de Ciencias Exactas y Naturales, Departamento de Física, Ciudad Universitaria, 1428 Buenos Aires, Argentina,}
\affiliation{CONICET - Universidad de Buenos Aires, Instituto de F\'{\i}sica Interdisciplinaria y Aplicada (INFINA), Ciudad Universitaria, 1428 Buenos Aires, Argentina.}
\author{S. Angriman}
\affiliation{Universidad de Buenos Aires, Facultad de Ciencias Exactas y Naturales, Departamento de Física, Ciudad Universitaria, 1428 Buenos Aires, Argentina,}
\affiliation{CONICET - Universidad de Buenos Aires, Instituto de F\'{\i}sica Interdisciplinaria y Aplicada (INFINA), Ciudad Universitaria, 1428 Buenos Aires, Argentina.}
\author{A. Ferran}
\affiliation{Université Grenoble Alpes, CNRS, Grenoble-INP, LEGI, F-38000 Grenoble, France}
\affiliation{Department of Mechanical Engineering, University of Washington, Seattle, Washington 98195-2600, USA}
\author{P. Cobelli}
\affiliation{Universidad de Buenos Aires, Facultad de Ciencias Exactas y Naturales, Departamento de Física, Ciudad Universitaria, 1428 Buenos Aires, Argentina,}
\affiliation{CONICET - Universidad de Buenos Aires, Instituto de F\'{\i}sica Interdisciplinaria y Aplicada (INFINA), Ciudad Universitaria, 1428 Buenos Aires, Argentina.}
\author{M. Obligado}
\affiliation{Université Grenoble Alpes, CNRS, Grenoble-INP, LEGI, F-38000 Grenoble, France}
\author{P.D. Mininni}
\affiliation{Universidad de Buenos Aires, Facultad de Ciencias Exactas y Naturales, Departamento de Física, Ciudad Universitaria, 1428 Buenos Aires, Argentina,}
\affiliation{CONICET - Universidad de Buenos Aires, Instituto de F\'{\i}sica Interdisciplinaria y Aplicada (INFINA), Ciudad Universitaria, 1428 Buenos Aires, Argentina.}

\begin{abstract} 
\end{abstract}

\maketitle 

\section*{Simulations of time-modulated HIT}

The CRC forcing generates a flow which, besides being unsteady, also has an inhomogeneous and anisotropic large-scale component. Thus, its extreme clustering could be influenced by these flow properties instead of being driven by turbulence unsteadiness. To help disentangle the effects of the large-scale flow from those caused by the modulation of the energy dissipation rate in particle clustering, we performed another DNS with a third type of forcing. In this DNS a random forcing was initially generated, exciting the same (isotropic) Fourier modes as those used in HIT. But instead of evolving the phases in time, the global amplitude of the forcing was modulated as
\begin{equation}
    F_0 (t) = A_0 \cos(f_0 t)^{\alpha},
    \label{eq:amp}
\end{equation}
where $A_0$ and $f_0$ are respectively the initial forcing amplitude, and the modulation frequency. Their values were chosen to have fluctuations in $\epsilon$ that resemble those in the CRC flow. The parameters used were $A_0 = 3.5 U_0^2/L_0$, $f_0 = (\pi/15) U_0/L_0$, and $\alpha = 100$. These values result in strong peaks in the forcing amplitude that repeat periodically, and that synthetically generate a modulation of $\epsilon$, $C_{\epsilon}$, and $R_{\lambda}$, but with a more homogeneous and isotropic forcing. We label this run as modulated HIT (mHIT) in the following. The simulation was made using a resolution of $512^3$ grid points and with a kinematic viscosity $\nu_\textrm{mHIT} = 1.1 \times 10^{-3} L_0 U_0$. As in all other simulations, this DNS has $\kappa \eta > 1$ at all times, where $\kappa = 512/3$, and a mean integral length scale $L\approx 1 L_0$.

\begin{figure} \centering
    \includegraphics[width=.5\linewidth]{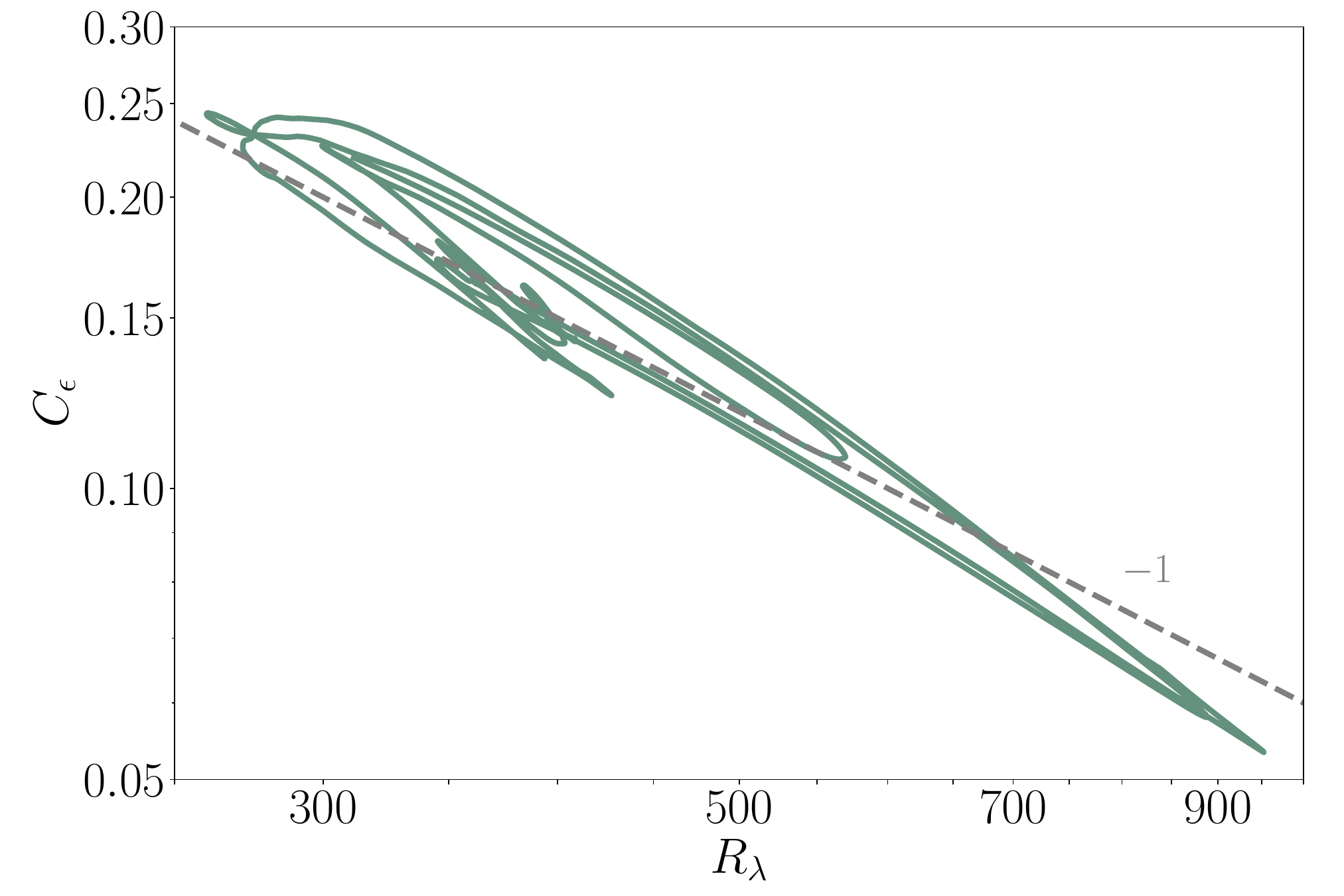} 
    \caption{Instantaneous value of $C_{\epsilon}$ as a function of $R_{\lambda}$ in the mHIT simulation. A slope of $-1$ is indicated as a reference.}
    \label{ceps_relam_sm}
\end{figure}

Figure \ref{ceps_relam_sm} shows the instantaneous value of $C_\epsilon$ as a function of $R_\lambda$ for mHIT. Note that the out-of-equilibrium dissipation law reported for other unsteady turbulent flows and found in the CRC flow is also compatible with this flow, with $C_{\epsilon} \sim \sqrt{\textrm{Re}_0}/R_{\lambda}$.

\begin{figure} \centering
    \includegraphics[width=.5\linewidth]{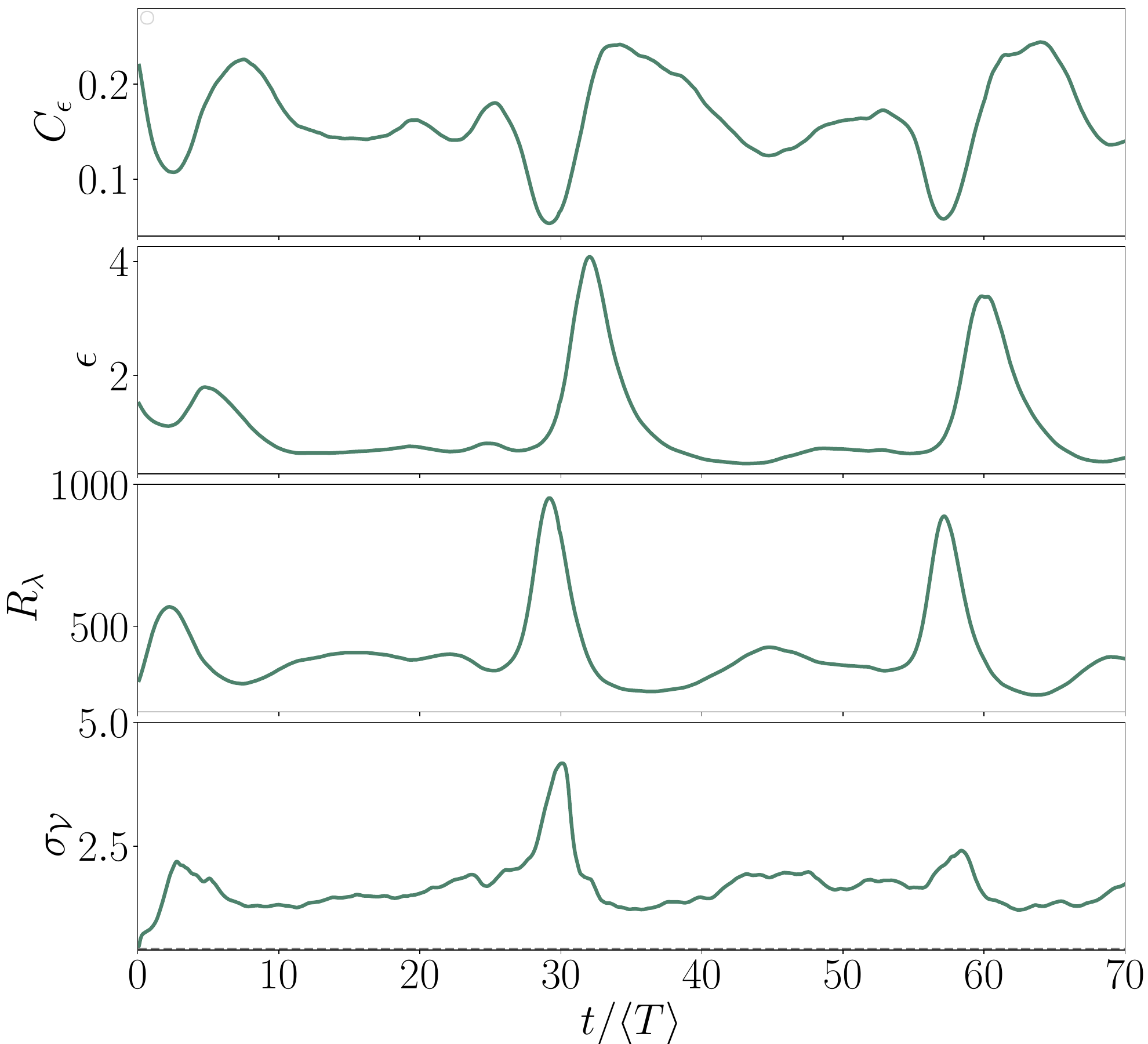} 
    \caption{Time series of (a) $C_{\epsilon}$, (b) $\epsilon$, (c) $R_{\lambda}$, and (d) $\sigma_{\altmathcal{V}}$ for mHIT with particles with $\textrm{St}=8$. Time is normalized by the mean turnover time $\left<T\right>$, and $\sigma_\textrm{RPP}$ is indicated as a reference by the dotted grey line in panel (d)}
    \label{ts_sm}
\end{figure}

\begin{figure} \centering
    \includegraphics[width=.5\linewidth]{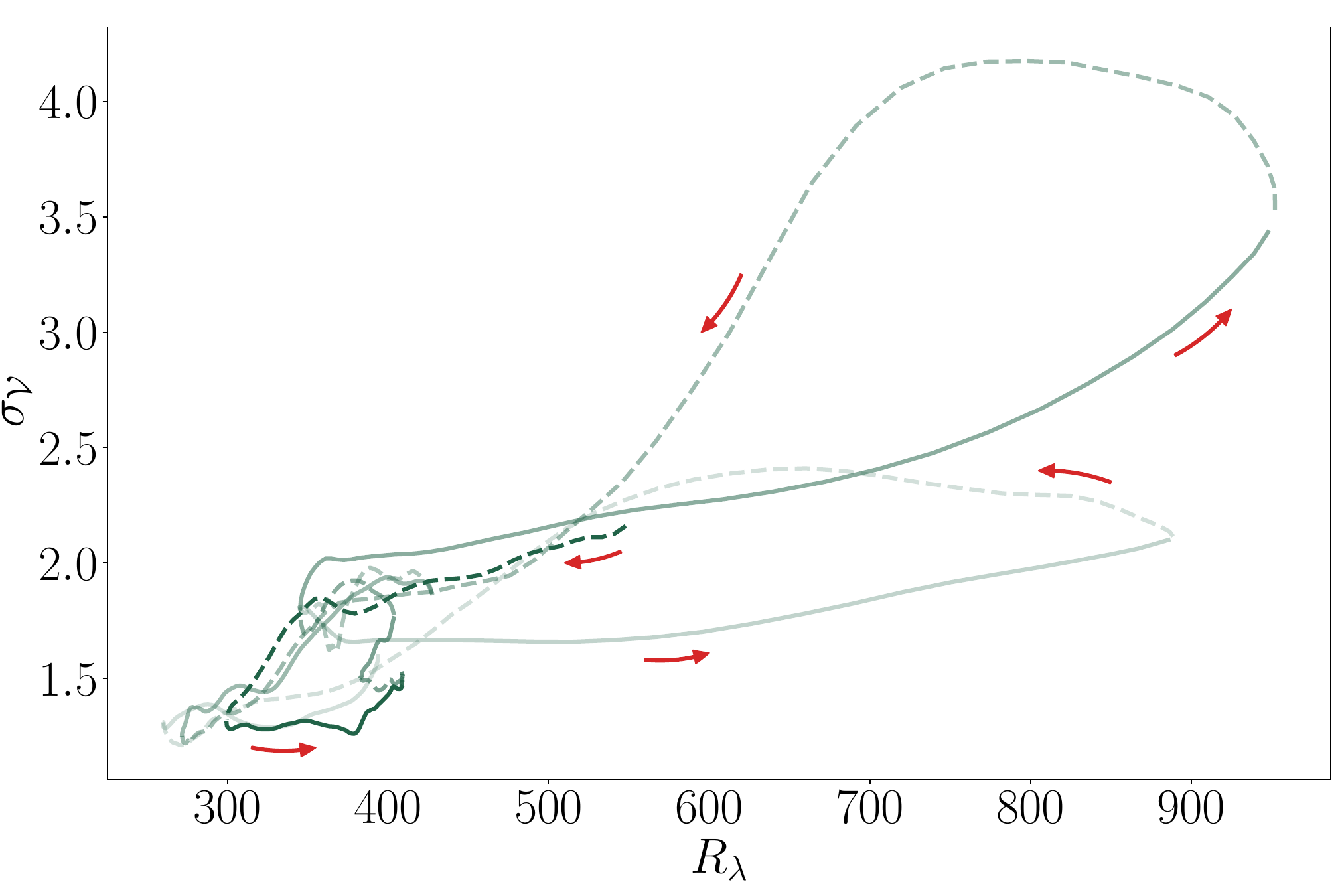} 
    \caption{Standard deviation of the Voronoi volumes, $\sigma_{\altmathcal{V}}$, as a function of $R_{\lambda}$ for the mHIT simulation, differenciating branches in which $R_{\lambda}$ increases (solid lines) or decreases (dashed lines). Red arrows indicate the direction of time evolution.}
    \label{hyst_sm}
\end{figure}

Following the same procedures as in the other simulations in this study, $10^6$ particles with $St = 8$ were injected when the flow reached a fully developed turbulent regime, at a time arbitrarily labeled as $t=0$, and at random positions in space. The instantaneous PDFs of the normalized Vorono\"i volumes were calculated for all particles positions at each time. Time series of $C_{\epsilon}$, $\epsilon$, $R_{\lambda}$, and $\sigma_{\altmathcal{V}}$ are shown in Fig.~\ref{ts_sm}. As in the other flows, in mHIT we also find that $\sigma_{\altmathcal{V}}$ is correlated (anticorrelated) with $R_{\lambda}$ ($C_{\epsilon}$ and $\epsilon$). Values for mHIT display strong fluctuations around a mean, as in the CRC flow. Fluctuations in $C_\epsilon$ and $\epsilon$ are directly caused by the amplitude modulation of the forcing, but $\sigma_{\altmathcal{V}}$ still responds to the fluctuations in the other quantities. The mean value of $\sigma_\nu$ is closer to that in HIT, but large excursions reaching $\sigma_\nu \gtrsim 4$ can be seen. This indicates that extreme clustering is mostly associated to the flow unsteadiness, while large-scale flow anisotropies and inhomogeneities can further increase particle clustering. Finally, Fig.~\ref{hyst_sm} shows $\sigma_{\altmathcal{V}}$ as a function of $R_{\lambda}$, confirming the presence of a hysteresis cycle and displaying a behavior similar to that observed in the CRC flow at different values of $R_{\lambda}$.

\section*{Time evolution of the CRC flow and of particles statistics}

A movie with particles positions and the time evolution of several quantities, for the CRC simulation with $512^3$ grid points and with particles with $\textrm{St} = 3$, is available as supplemental material. The top panel shows $\sigma_\altmathcal{V}$ and $C_\epsilon$ as a function of time, the bottom left panel shows instantaneous probability density functions (PDFs) of the normalized Vorono\"i volumes (a random Poisson process is indicated as a reference by the dashed line), and the top right panel shows particles in the $xz$ plane for a slice of the box centered around $y = (\pi/2) L_0$. Note the accumulation of particles when $\sigma_\altmathcal{V}$ becomes larger, and the more homogeneous distribution when $\sigma_\altmathcal{V}$ becomes smaller.

%\bibliographystyle{apsrev4-2}
%\bibliography{ms}